\RequirePackage{fix-cm}
\documentclass[smallextended]{svjour3}       
\smartqed  
\usepackage{xcolor}
\usepackage{graphicx}
\usepackage{amsmath,amssymb}
\usepackage{subfig}
 \journalname{Gen. Rel. Grav.}
\begin{document}

\title{Gravitational Wave Signatures of Highly Magnetized Neutron Stars
}


\author{Cesar V. Flores        \and
        Luiz L. Lopes \and
        Luis B. Castro \and
        D\'ebora P. Menezes 
}


\institute{Cesar V. Flores \at
              Departamento de F\'isica, Universidade Federal de Santa Catarina, Florian\'opolis, SC, CP 476, CEP 88.040-900, Brazil. \\
              \email{cesarovfsky@gmail.com}           
           \and
           Luiz L. Lopes \at
              Centro Federal de Educa\c{c}\~ao Tecnol\'ogica de Minas Gerais, Campus VIII; CEP 37.022-560, Varginha, MG, Brasil. \\
              \email{llopes@varginha.cefetmg.br}           
            \and
           Luis B. Castro \at
           Departamento de F\'isica - CCET,  Universidade Federal de Maranh\~ao, Campus Universit\'ario de Bacanga, S\~ao Lu\'is, MA, CEP 65080-805, Brazil. \\
           \email{luis.castro@pq.cnpq.br,\,lrb.castro@ufma.br}
           \and
           D\'ebora P. Menezes \at
           Departamento de F\'isica, Universidade Federal de Santa Catarina, Florian\'opolis, SC, CP 476, CEP 88.040-900, Brazil. \\
              \email{debora.p.m@ufsc.br}           
}

\date{Received: date / Accepted: date}

\maketitle

\begin{abstract}
Motivated by the recent gravitational wave detection by the LIGO-VIRGO observatories, we study the Love number and dimensionless tidal polarizability  of highly magnetized stars. We also investigate the fundamental quasi-normal mode of neutron stars subject to high magnetic fields. To perform our calculations we use the chaotic field approximation and consider both nucleonic and hyperonic stars. As far as the fundamental mode is concerned, we conclude that the role played by the constitution of the stars is far more relevant than the intensity of the magnetic field and if massive stars are considered, the ones constituted by nucleons only present frequencies somewhat lower than the ones with hyperonic cores, a feature that can be used to point out the real internal structure of neutron stars. Moreover, our studies clearly indicate that strong magnetic fields play a crucial role in the deformability of low mass
neutron stars, with possible consequences on the interpretation of the detected gravitational waves signatures.
\keywords{Gravitational waves \and magnetized neutron star \and non-radial oscillations}
\end{abstract}

\section{Introduction}
\label{intro}
On August 17, 2017 the LIGO-VIRGO collaboration observed the gravitational wave event GW170817, which consisted in the detection of a binary neutron star merger and, as consequence established a new channel to study the high density equation of state (EOS) that describes neutron stars. 
In these kind of events, before the merging, the neutron star components begin to react to their mutual tidal fields, and this effect can be detected in the phase modification  of the gravitational wave impinging on the detector. This tidal response depends strongly on the neutron star composition  and therefore important information can be obtained about the EOS \cite{PhysRevD.77.021502,PhysRevD.79.124033}. In addition another source of information can be inferred from  neutron star oscillations. In principle, before the merging, tidal interactions can excite the fluid modes by resonance
\cite{shibata1994,parisi2017} and also during and after the neutron star fusion, the fundamental mode can be greatly excited, with a strong influence on the respective gravitational wave emission \cite{bauswein2016,andersson2011,faber2012,rezzolla2013}. Besides gravitational waves, the associated IR, optical, UV, x-ray and $\gamma$-ray electromagnetic
radiations were also detected \cite{SoaresSantos2017}, giving rise to the multimessenger astronomy era. 
Five years before the first detection of a neutron star merger, the possibility that the
resonant excitation of neutron star modes by tides could also be seen as a source of short gamma-ray bursts had already been proposed \cite{Tsang2012} as a complementary way to probe the neutron star structure.

At present, a trustworthy determination of the EOS of strongly interacting matter at densities above few times the nuclear saturation density remains a challenge. The EOS can be reliably obtained up to the nuclear saturation density, but at densities typical of neutron star interiors, which can be up to 6 times higher, its determination depends crucially on the knowledge of strong interactions in a regime that cannot be reached on earthly experiments. For this reason, the validity of both non-relativistic (Skyrme-type) and relativistic (RMF) models has to be checked according to verifiable constraints. In the present work we restrict our comments and calculations to relativistic models.

Not too long ago, 263 RMF models were confronted with nuclear bulk properties inferred from experiments. Three different sets of constraints were built in accordance with largely accepted properties related to compressibility, pressure, symmetry energy and its derivatives. Although all those models were originally proposed to describe specific nuclear physics quantities, when analysed with respect to these general sets of constraints, only 35 of them satisfied all of them \cite{PhysRevD.77.021502}. These 35 models were then investigated with respect to
stellar matter properties in \cite{PhysRevC.93.025806} and only 12 parametrizations resulted in neutron stars with a maximum mass in the range of 1.93-2.05 $M_\odot$, as the ones measured in the present decade \cite{demorest2010,antoniadis2013}. 
Recently, an even more massive pulsar was detected, the MSP J0740+6620, which has a mass of $2.14^{+0.10}_{-0.09}$ at 68$\%$ credibility interval and $2.14^{+0.20}_{-0.18}$ at 95$\%$  credibility interval~\cite{Cromartie}. If it turns out to be above 2.1 $M_{\odot}$, another mechanism may be necessary to account for this pulsar and for similar ones, not detected yet.

As far as the constraints imposed by GW170817 are concerned,
34 out of the 35 models were also confronted with  them \cite{PhysRevC.99.045202} (we come back to these constraints later in the text) and 24 were shown to satisfy these constraints. Nevertheless, only 5 RMF parametrizations can simultaneously describe massive stars and GW170817 constraints. If hyperons are included in the calculations, the situation becomes even more complicated because the EOS must be soft at subsaturation densities and hard at higher densities to predict massive stars, but hyperons soften the EOS. There are different ways to circumvent this problem and one of them is by introducing the magnetic field in the Lagrangian density that describes the model, a field which is capable of stiffening the EOS.  

However, the main reason to consider magnetic field effects on the EOS is the existence of magnetars, which are a special class of neutron stars bearing surface magnetic fields that are three orders of magnitude stronger than the ones present in their non-magnetized counterparts \cite{2015SSRv..191..315M} ($10^{12}$ G). So far, only 30 of them have been clearly identified \cite{2014ApJS..212....6O}, but the launching of NICER \cite{NICER} in 2017 and ATHENA \cite{ATHENA}, expected to take place in 2030, will certainly provide more information on these compact objects. Moreover, most of the known magnetars detected so far as either transient X-ray sources, known as soft-gamma repeaters or persistent anomalous x-ray pulsars, are isolated objects. Although, possible manifestations of accreating magnetars have been found \cite{Tong2018},
all the analyses done in the present work in relation with GW170817 constraints, coming from a binary system, have to be taken with care.

At this point, it is important to mention that there is some controversy on how strong magnetic fields can be incorporated to the Lagrangian density and the stress tensor. While some authors advocate that the EOS should be isotropic and no magnetization could appear in the EOS \cite{scalarp,PhysRevC.85.039801}, others claim that the anisotropy is indeed present and magnetization effects should be considered \cite{PhysRevD.83.043009,PhysRevD.98.043022,PhysRevC.91.065205}. For a discussion on this subject, the reader can refer to \cite{LLL2016}, where the different results obtained with the two formalisms can be seen. There is no doubt that the ideal situation is to use the LORENE code \cite{LORENE}, which performs a numerical computation of the neutron star by taking into account Einstein-Maxwell equations and equilibrium solutions self consistently with a density dependent magnetic field. Unfortunately, this calculation is not always feasible for all purposes and it gives some estimates that may not be correct, as the case of the neutron star crust thickness  discussed in \cite{Chatterjee2019}. However, the latter work shows clearly that although not too strong magnetic fields have negligible effect on the EOS itself, it strongly affects properties related to the crust, as the cases discussed next. 

Another feature worth investigating in magnetars is the possible oscillations they can produce. These oscillations result in quasi-normal modes and a family of modes exists, which offers a great opportunity to test the gravitational wave asteroseismology approach in neutron stars.  
Future  third  generation  detectors,  like  the  Einstein Telescope \cite{Hild08},  will  have enough sensitivity  to  observe  the quasi-normal  modes  of compact objects.
From the point of view of detectability, the most promising modes are the crustal modes, $g$-modes, $f$-modes and $r$-modes. In this work we focus only on the $f$-mode because it is more easily excited and it is expected to be the first one to be detected.

In the present paper, we use the chaotic field approximation introduced by Zel'dovich \cite{Zeldovich} and applied in \cite{LLL2015,xing,LLL2016,LLL2020} to account for strong magnetic fields in NS. The EOS is calculated for nuclear matter and hyperonic matter and both situations are investigated in the context of non-radial oscillations and tidal polarizabilities.

The paper is organized as follows: in Sect. II we review some basic aspects of the EOS. In Sect. III, the formalism used to compute tidal deformability and related quantities is resumed and in Sect. IV 
we present the equilibrium configuration and the non-radial oscillation equations for neutron stars. Along these three sections the results are displayed and discussed.
In Sect. V we make our final remarks. 

\section{Magnetized Equation of state} \label{EOS}
If we believe that the standard model is correct, the physics of strong interacting matter is described by quantum chromodynamics (QCD). However, QCD provides no meaningful results in the  region of the neutron star interior, i.e., high density and low temperature. To overcome this issue, we use an effective model, the quantum hadrodynamics (QHD). Originally developed in the early 70s~\cite{Walecka:1974qa}, QHD considers the baryons, not the quarks, as the fundamental degrees of freedom. Also, the strong interaction is simulated by the exchange of massive mesons. In this work we use an extended version of the QHD whose lagrangian density reads~\cite{Serot}:
\begin{eqnarray}
\mathcal{L}_{QHD} = \sum_b \bar{\psi}_b \bigg [\gamma^\mu(i\partial_\mu -e_bA_\mu - g_{b v}\omega_\mu  - g_{b \rho} \frac{1}{2}\vec{\tau} \cdot \vec{\rho}_\mu) 
- (m_b - g_{b s}\sigma ) \bigg ]\psi_b  \nonumber \\   + \frac{1}{2} m_v^2 \omega_\mu \omega^\mu 
    + \frac{1}{2} m_\rho^2 \vec{\rho}_\mu \cdot \vec{\rho}^{ \; \mu}   + \frac{1}{2}(\partial_\mu \sigma \partial^\mu \sigma - m_s^2\sigma^2)  
    - U(\sigma) \nonumber \\ - \frac{1}{4}F^{\mu \nu}F_{\mu \nu}  - \frac{1}{4}\Omega^{\mu \nu}\Omega_{\mu \nu} -  \frac{1}{4}\bf{P}^{\mu \nu} \cdot \bf{P}_{\mu \nu}  , \label{s1} 
\end{eqnarray}
in natural units. The sum in $b$ stands just for the nucleons or for all the baryon octet, depending on our choice for the star constituents,
 $\psi_b$  are the    Dirac fields of the baryons,  $\sigma$, $\omega_\mu$ and $\vec{\rho}_\mu$ are the mesonic fields,
 and $A_\mu$ is the electromagnetic four-potential.
 The $g's$ are the Yukawa coupling constants that simulate the strong interaction, $m_b$ and $e_b$  are the mass and the electric
charge  of the baryon $b$; $m_s$, $m_v$, and $m_\rho$ are
 the masses of the $\sigma$, $\omega$,  and $\rho$ mesons respectively.
 The antisymmetric  field  tensors are given by their usual expressions as presented in~\cite{Glendenning-book}.
  The $U(\sigma)$ is the self-interaction term introduced in Ref.~\cite{Boguta} to fix some of the saturation properties of the nuclear matter.
We also define $M^{*}_b$ as the effective mass of the baryon $b$: $M^{*}_b = M_b - g_{b s}\sigma$.

In the presence of a magnetic field $B$ in the $z$ direction, the energy eigenvalue $E_b$,
and the number density $n_b$ of charged baryons are quantized:

\begin{equation}
E_b = \sqrt{M^{*2}_b - k_z^2 + 2s|e|B} , \quad n_b = \sum_\nu \frac{|e|B}{2\pi^2} k_z \label{s2},
\end{equation}
where the discrete parameter $s$ is called Landau level $(LL)$. The uncharged baryon energies are not modified by the magnetic interaction and keep their 
usual expressions~\cite{Glendenning-book}. The mesonic fields are obtained by mean
field approximation~\cite{Serot,Glendenning-book,LLL2012} 
and the equation of state (EOS)  by thermodynamic
relations~\cite{Greiner}. To construct a 
$\beta$ stable matter, we also 
include leptons as a free Fermi gas and impose zero net charge and chemical equilibrium.

To describe the properties of nuclear matter, we use a slightly modified version of the well-known GM1 parametrization \cite{glendenning1991}, which is 
a widely accepted parametrization \cite{Weiss1,Weiss2} that is able to reasonably describe both, nuclear matter and stellar structure, consistent with experimental and
astrophysical observations \cite{LLL2013}. Here, we just reduce the strength of the $\rho$ coupling, reducing the symmetry energy slope $L$ from the original 94 MeV to 87.9 MeV \cite{LLL2014}, a value closer to what is inferred in recent observations \cite{Steiner1,Steiner2}. 
We expect that the same qualitative behavior of magnetized neutron stars 
is obtained with any other parametrization.

In Table \ref{TL1} we show the parameters of the model and its previsions for five nuclear matter properties at saturation density: saturation density
point ($n_0$), incompressibility ($K$), binding energy per baryon ($B/A$), symmetry energy ($S_0$) and its slope ($L$). 

\begin{table}[ht]
\begin{center}
\caption{Slightly modified GM1 parametrization. Parameters of the model and nuclear bulk property previsions. $N$ represents both nucleons.} 
\label{TL1}
\begin{tabular}{|c|c||c|c|c|}
\hline
 \multicolumn{2}{|c||}{ Parameters} & \multicolumn{2}{c|}{Previsions at $n_0$} \\

 \hline
 $(g_{N\omega}/m_v)^2$   &  7.148 $fm^2$  & $n_0~(fm^{-3})$  &  0.153     \\
 \hline
  $(g_{N\sigma}/m_s)^2$ & 11.785 $fm^2$  & $K$ (MeV) & 300    \\
 \hline
   $(g_{N\rho}/m_\rho)^2$ & 3.880 $fm^2$  & $B/A$ (MeV) & -16.3     \\
 \hline
  $\kappa/M_N$  & 0.005894 & $S_0$ (MeV)& 30.5  \\
\hline
 $\lambda$ & -0.006426 & $L$ (MeV) & 87.9    \\
\hline
\end{tabular} 
\end{center}
\end{table}

Now, we discuss the presence (or the absence) of hyperons in the neutron stars core. The possibility of the hyperon onset in neutron stars is an old~\cite{old} but yet, very active subjetc of study~\cite{vidana2016}. This is the so called hyperon puzzle. The main problem is that we have very little knowledge of how the hyperons interact with nucleons and with each other, or in QHD words, what are the hyperon-meson coupling constants? Since we have six hyperons and three mesons, we have, in principle eighteen free parameters besides those presented in Table~\ref{TL1}. To overcome these 
profusion of parameters, we rely on symmetry group techniques. Following Ref.~\cite{LLL2013}, we use the hybrid symmetry group SU(6) to fix all hyperon-vector meson coupling constants and a $nearly$ SU(6)  symmetry to fix the hyperon-scalar meson coupling constants.  This reduces the eighteen free parameters to just one, which in turn, is fixed using the $\Lambda$ hyperon potential depth: $U_\Lambda = - 28$ MeV.  The values we obtain are:

\begin{eqnarray}
\frac{g_{\Lambda\omega}}{g_{N\omega}} = \frac{g_{\Sigma\omega}}{g_{N\omega}} = 0.667,  \quad \frac{g_{\Xi\omega}}{g_{N\omega}} =  0.333, \nonumber \\
\frac{g_{\Sigma\rho}}{g_{N\rho}} = 2.0 \quad \frac{g_{\Xi\rho}}{g_{N\rho}} = 1.0 , \quad \frac{g_{\Lambda\rho}}{g_{N\rho}} = 0.0,  \\ \label{EL1}
\frac{g_{\Lambda\sigma}}{g_{N\sigma}} = 0.610 , \quad \frac{g_{\Sigma\sigma}}{g_{N\sigma}} =  0.396 , \quad \frac{g_{\Xi\sigma}}{g_{N\sigma}} = 0.113 .\nonumber
\end{eqnarray}

Before finishing this section, we discuss the influence of the magnetic field itself on the EOS. As pointed out earlier, the ideal situation would be to use the LORENE code \cite{LORENE}, but its use is not possible to compute all desirable quantities investigated in the present work. Hence, here we use an alternative, the so called chaotic magnetic field approximation. As presented by Zel'dovich~\cite{Zeldovich}, we can only use the concept of pressure, when we are dealing with a small-scale chaotic field. In this case, the stress tensor reads ${\rm diag}=(B^2/6, B^2/6, B^2/6)$. The chaotic magnetic field approximation, as pointed in ref.~\cite{LLL2015} has the advantage of restoring the thermodynamic consistency of the model as it deals with the scalar concept of   pressure \cite{scalarp}. Also, when compared with results obtained with the LORENE~\cite{LORENE} code, we see that the chaotic magnetic field does not overestimate the maximum mass either, in contrast with other prescriptions. We comment next on our results and the validity of the chaotic magnetic field approximation in the light of a study published in Ref.~\cite{debi2018} 
concerning the magnetic field distribution in magnetar interiors. Using a full relativistic numerical calculation, it was found that the magnetic field can be expressed as a multipolar expansion that accounts for the monopole contribution, the dipole term, the quadrupole and so on. Now, the important fact here is is that the chaotic magnetic field formalism  is the monopole approximation for the magnetic field profile. As shown in Fig 3 of Ref.~\cite{debi2018}, the monopole term is dominant throughout almost the entire star. More than that, the monopole term is specially dominant in the neutron star core, 
when the magnetic field is stronger. So, in the limit of a very strong field, when its influence is bigger, our results are very close to those obtained with a more sophisticated formalism. Moreover, one of the main problems of using the TOV equations in the presence of strong magnetic fields is the possible appearance of anisotropies in the 
momentum-energy tensor. As pointed out in Ref.~\cite{debi2018} in most cases, $T_{\theta \theta} \neq T_{rr}$. However, exactly due to the monopole nature of the chaotic
magnetic field, we always obtain $T_{\theta \theta} = T_{rr}$ , which guarantees that the TOV approximation can be used in this case.

We also briefly discuss the limitations of the chaotic magnetic field approximation: as a truly isotropic, spherically symmetric approximation, anisotropies in both neutron star structure ~\cite{Gomes1,Gomes2,Gomes3} and in the microscopic EOS ~\cite{Efrain1,Efrain2} are beyond the scope of this manuscript. Notice however, that to obtain significant deformation on the neutron stars ~\cite{Gomes1,Gomes3}, the use of unrealistic fields up to 
$ 3 \times 10^{17}$ G at the surface seems to be necessary.
These values can be literally a thousand times stronger than what is observed in magnetars. Moreover, a density dependent magnetic field violates Maxwell equations as discussed in ~\cite{alloy16} and a rearrangement term, never calculated, would be necessary.

In our case, the EOS reads:

\begin{equation}
\epsilon_T =  \epsilon_M + \frac{B^2}{2}  , \quad P_T = P_M + \frac{B^2}{6}.  \label{s7}
\end{equation}
where the subscript $M$ stands for the matter contribution to the EOS.

As mentioned in the Introduction, magnetars bear a magnetic field of the order of $10^{15}$ G at the surface but according to the Virial theorem, stronger fields can be expected in their interior.
To account for the growing of the magnetic field strength towards the neutron star core, we follow Ref.~\cite{LLL2015,LLL2016,xing} and use an energy density-dependent magnetic field:

\begin{equation}
B =   B_0 \bigg ({\frac{\epsilon_M}{\epsilon_c}} \bigg )^{\alpha} + B^{surf}, \label{EL5}
\end{equation}
where  $\epsilon_c$  is the energy density at the center of the maximum mass neutron star with zero magnetic field and $\alpha$ is any positive number, reducing the number of free parameters from two to only one. Moreover, as explained in detail in Ref. \cite{LLL2015}, if we take $\alpha > 2$, the model becomes practically parameter free. In eq. (\ref{EL5}), $B_0$ is then fixed value of the magnetic field that is taken next as $1.0 \times 10^{18}G$, $3.0 \times 10^{18}G$, or zero. With this recipe,
the magnetic field is no longer fixed for all neutron star configurations.
Each EOS produces a different value for $\epsilon_c$ that enters in Eq. (\ref{EL5}). For our particular case, $\epsilon_c = 4.98 fm^{-4}$ for neutron stars with hyperons and $\epsilon_c = 5.65 fm^{-4}$ for neutron stars without hyperons in the core, ensuring  that the magnetic field does not exceed  $B_0$. In this work we use $\alpha = 3.$

\begin{figure}
\centering
\includegraphics[angle=0,scale=0.35]{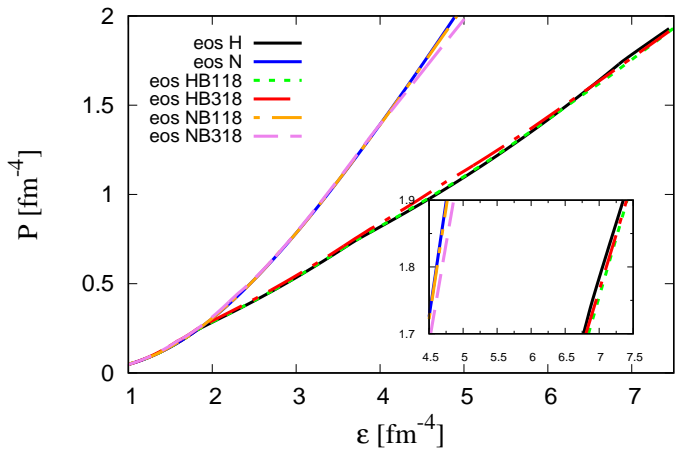}
\caption{EOS obtained with nucleons only (labelled N) and with the baryon octet (labelled H) for non-magnetized matter and with $B_0$ equal to  $1.0 \times 10^{18}G$ (labelled B118) and $3.0 \times 10^{18}G$ (labelled B318).}
\label{fig1}
\end{figure}

Before we proceed, we show in Fig. \ref{fig1} all the EOS that we use in the present work. As it is always the case, hyperons soften the EOS and magnetic fields within the chaotic approximation change the EOS only slightly, generally making them a bit stiffer.
We can see two groups of curves, each one with three EOS, the softer ones describing matter with hyperons and the stiffer one, nucleonic matter.

\section{Tidal deformability and  GW170817 constraints}

\begin{figure}
\includegraphics[angle=-90,scale=0.35]{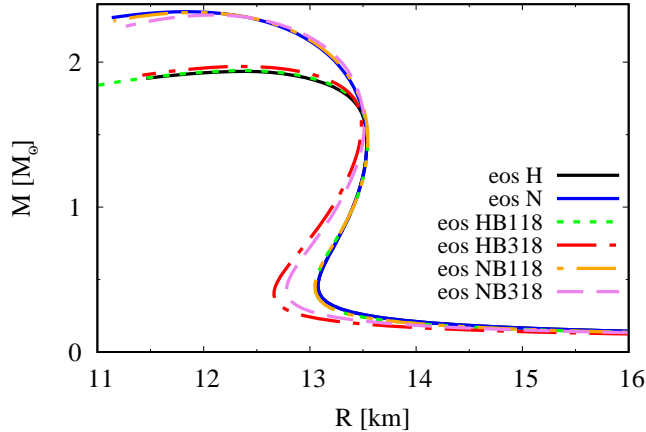} 
\caption{ The mass-radius relation is plotted for the six EOS shown in Fig. \ref{fig2}.}
\label{fig2}
\end{figure}

To study the tidal deformability, at first we have to solve the equilibrium configuration for the neutron star, which is represented by the  Tolman-Oppenheimer-Volkoff (TOV) equations given by
\begin{eqnarray}
\label{tov1}
\frac{dp}{dr} &=& - \frac{\epsilon m}{r^2}\bigg(1 + \frac{p}{\epsilon}\bigg)
	\bigg(1 + \frac{4\pi p r^3}{m}\bigg)\left(1 - \frac{2m}{r}\right)^{-1},  \\
\label{tov2}
\frac{d\nu}{dr} &=& - \frac{2}{\epsilon} \frac{dp}{dr} \bigg(1 + \frac{p}{\epsilon}\bigg)^{-1}, \\
\label{tov3}
\frac{dm}{dr}& =& 4 \pi r^2 \epsilon,
\end{eqnarray}
where  $m$ and $\nu$ are respectively the gravitational mass and the metric potential. The pressure $p$ and the mass-energy density $\epsilon$ are given by the equations of state given in section II. The initial conditions at the centre are $m(0) =0$, $p(0)=p_c$ and $\nu(0) = \nu_c$. We stop the integration when the pressure becomes zero, and at this point we define the surface of the star whose radius is $R$. We also apply the junction condition to the metric $\nu(R)= \ln ( 1- {2M}/{R} )$. 
We next use the EOS described in the last section as input to the TOV equations~\cite{tov} to obtain the macroscopic properties of the neutron stars. The mass-radius relations are plotted in Fig.~\ref{fig2} and the main properties are displayed in Table~\ref{TL2}.
 
\begin{table}[ht]
\begin{center}
\caption{Neutron stars main properties for each one of the six EOS. Indicating the maximum mass, the respective radius, central density, the radius of the canonical 1.4$M_\odot$ star and the central magnetic field for the maximum mass and the canonical one.} 
\label{TL2}
\begin{tabular}{|c|c|c|c|c|c|c|c|}
\hline
  EOS   &  $M_{max}$ ($M_{\odot}$) & R $(km)$  &  $\epsilon_{c}$ ($ fm^{-4}$) & $B_c$  (G) & $R_{1.4} (km) $  
  & $B_{c(1.4)}$  (G)\\
 \hline
 H & 1.95   & 12.46 & 4.98   & - &  13.63  & -\\
 \hline
  N & 2.37   & 11.91 & 5.65  &  - &13.63  & - \\
 \hline
 HB118  & 1.95 & 12.47 & 4.96 &  8.8 x 10$^{17}$ &13.64 & 3.0 x 10$^{16}$ \\
\hline
 HB318 & 1.98 & 12.47 & 4.94 & 2.7 x 10$^{18}$  & 13.58  & 1.0 x 10$^{17}$\\
\hline
 NB118 & 2.36 & 11.96 & 5.60  &  8.8 x 10$^{17}$ & 13.64 &  2.3 x 10$^{16}$ \\
\hline
 NB318 & 2.36 & 12.10 & 5.45  & 2.6 x 10$^{18}$ & 13.58   &  7.4 x 10$^{16}$  \\
\hline
\end{tabular} 
\end{center}
\end{table}

 We see that the effect of the chaotic magnetic field is to increase a little the maximum mass of stars with hyperonic core, but causes no significant variation on the neutron stars without hyperons. This small increase on the maximum mass is in agreement with results obtained with the LORENE code \cite{LORENE}. We can also see that the canonical mass star bears a magnetic field that is not too high, never surpassing 1 x 10$^{17}$ G. Such value agrees with estimates for realistic  stable  canonical mass stars~\cite{Reis},
 although in the very early stages of a magnetar life it may exceed this value through dynamo activity, Kelvin-Helmholtz or MRI instabilities~\cite{Price}. We can also see from Fig. \ref{fig2} that the low mass neutron stars with magnetic field bear a slightly smaller radii when compared with the stars without magnetic field. For a strong magnetic field ($B_0$ =  $3 \times 10^{18}G)$ the radii of the canonical stars (the ones with 1.4 $M_\odot$) are always smaller.  We can also see that with the GM1 model, even with hyperons, the maximum masses are very close to 2.0 solar masses, a necessary constraint since the observations of massive NS \cite{demorest2010,antoniadis2013,Cromartie}.

After the computation of the equilibrium configuration we proceed to the study of tidal deformations, which depend on the internal structure of neutron stars and our purpose is to use it to constraint the equation of state of the magnetars. For this objective we present, in the next lines, some comments about the theory of tidal deformabilities, the main equations and relationships that are necessary for our calculations. 

The relativistic theory of tidal effects was deduced by Damour and Nagar, Binnington and Poisson \cite{PhysRevD.80.084035,PhysRevD.80.084018}. They concluded that the tidal deformation of a neutron star is characterized by the gravito-electric $K^{el}_{2}$ and gravito-magnetic $K^{mag}_{2}$ Love numbers, where the former is related to the mass quadrupole and the second to the current quadrupole induced by the companion star. Further researches by Flanagan and Hindeler concluded that only a single detection should be sufficient to impose upper limits on $K^{el}_{2}$ at 90\% confidence level
\cite{PhysRevD.77.021502}. Since then intense research has been invested on the computing of Love numbers of neutron stars \cite{PhysRevC.87.015806,PhysRevC.98.035804,PhysRevC.98.065804,PhysRevC.95.015801,PhysRevD.81.123016,Hinderer_2008}.  

In a binary system the induced quadrupole moment $Q_{ij}$ in one neutron star due to the external tidal field ${\cal E}_{ij}$ created by a companion compact object can be written as~\cite{PhysRevD.81.123016,Hinderer_2008},
\begin {equation}
Q_{ij} = -\lambda {\cal E}_{ij},
\end{equation}
where, $\lambda$ is the tidal deformability parameter, which can be expressed in terms of dimensionless $l = 2$ quadrupole tidal Love number $k_2$ as
\begin{equation}
\label{tidal}
\lambda= \frac{2}{3} {k_2}R^{5}.
\end{equation}
To obtain $k_{2}$ we have to solve the following differential equation  
\begin{equation}
r \frac{dy}{dr} + y^2 + y F(r) + r^2Q(r)=0,
\label{ydef}
\end{equation}
where the coefficients  are given by
\begin{equation}
F(r) = [1 - 4\pi r^2(\varepsilon - p)]/E
\end{equation}
and
\begin{align}
Q(r)&=4\pi \left[5\varepsilon + 9p + (\varepsilon + p)\left(\frac{\partial 
p}{\partial\varepsilon}\right)-\frac{6}{4\pi r^2}\right]/E 
\nonumber\\ 
&- 4\left[ \frac{m+4\pi r^3 p}{r^2 E} \right]^2,
\end{align}
with $E = 1-2m/r$, $\varepsilon$ and $p$ are the energy density and pressure profiles inside the star. Therefore the Love number $k_2$ can be obtained as
\begin{align}
&k_2 =\frac{8C^5}{5}(1-2C)^2[2+C(y_R-1)-y_R]\times
\nonumber\\
&\times\Big\{2C [6-3y_R+3C(5y_R-8)]
\nonumber\\
&+4C^3[13-11y_R+C(3y_R-2) + 2C^2(1+y_R)]
\nonumber\\
&+3(1-2C^2)[2-y_R+2C(y_R-1)]{\rm ln}(1-2C)\Big\}^{-1},
\label{k2}
\end{align}
where $y_R = y(r = R)$  and $C= M/R$ are the star compactness, $M$ and $R$ are the total mass and radius of the star respectively.  Equation (\ref{ydef}) has to be solved coupled to the TOV equations.

The dimensionless tidal deformability $\Lambda$ (i.e., the dimensionless version of $\lambda$) is connected with the compactness parameter $C$ through 
\begin{equation}
\Lambda= \frac{2k_2}{3C^5}.
\label{dtidal}
\end{equation}  

\begin{figure}[ht]
\subfloat[]{\includegraphics[angle=-90,width=0.5\textwidth]{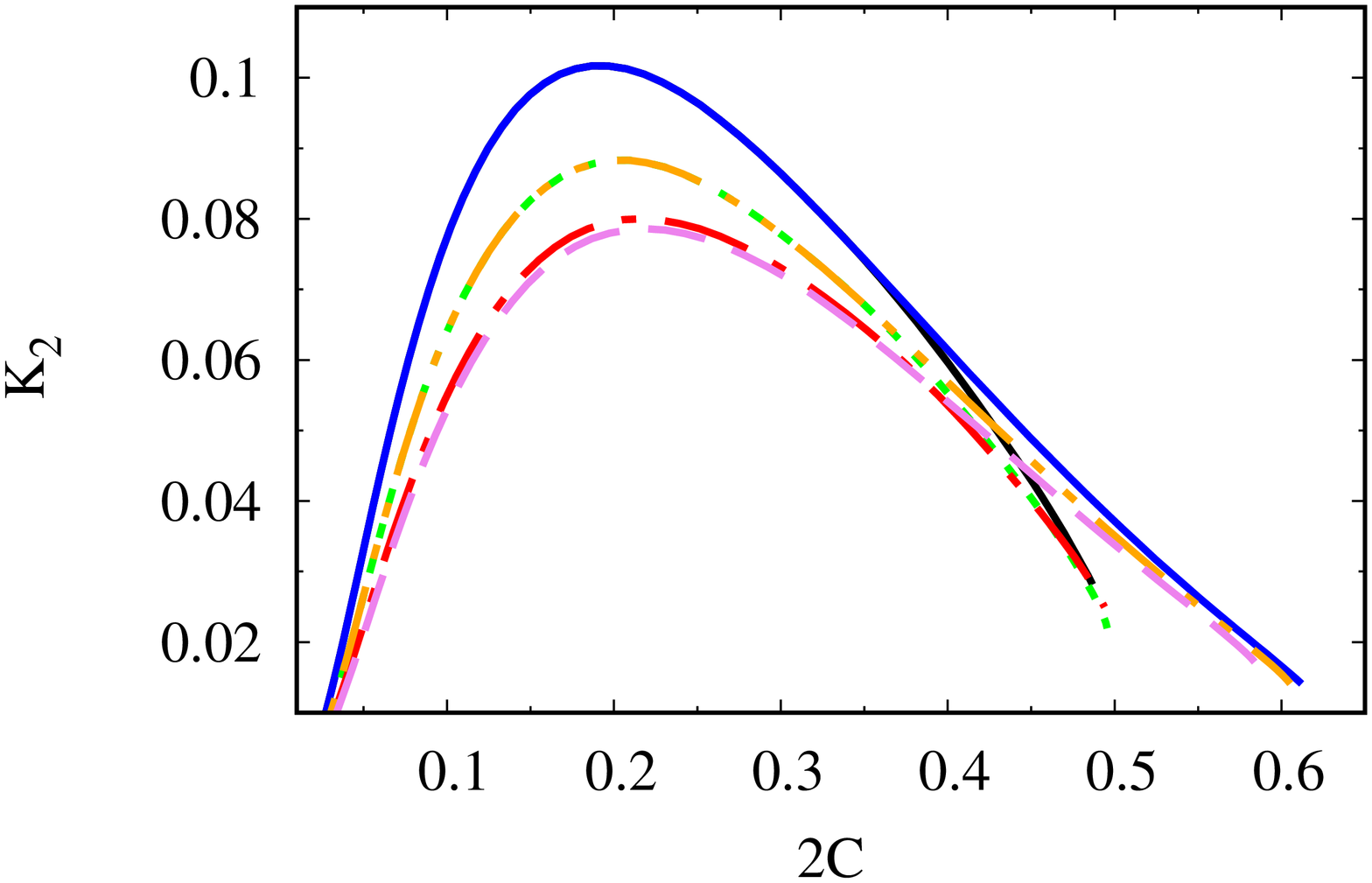}}
\quad
\subfloat[]{\includegraphics[angle=-90,width=0.5\textwidth]{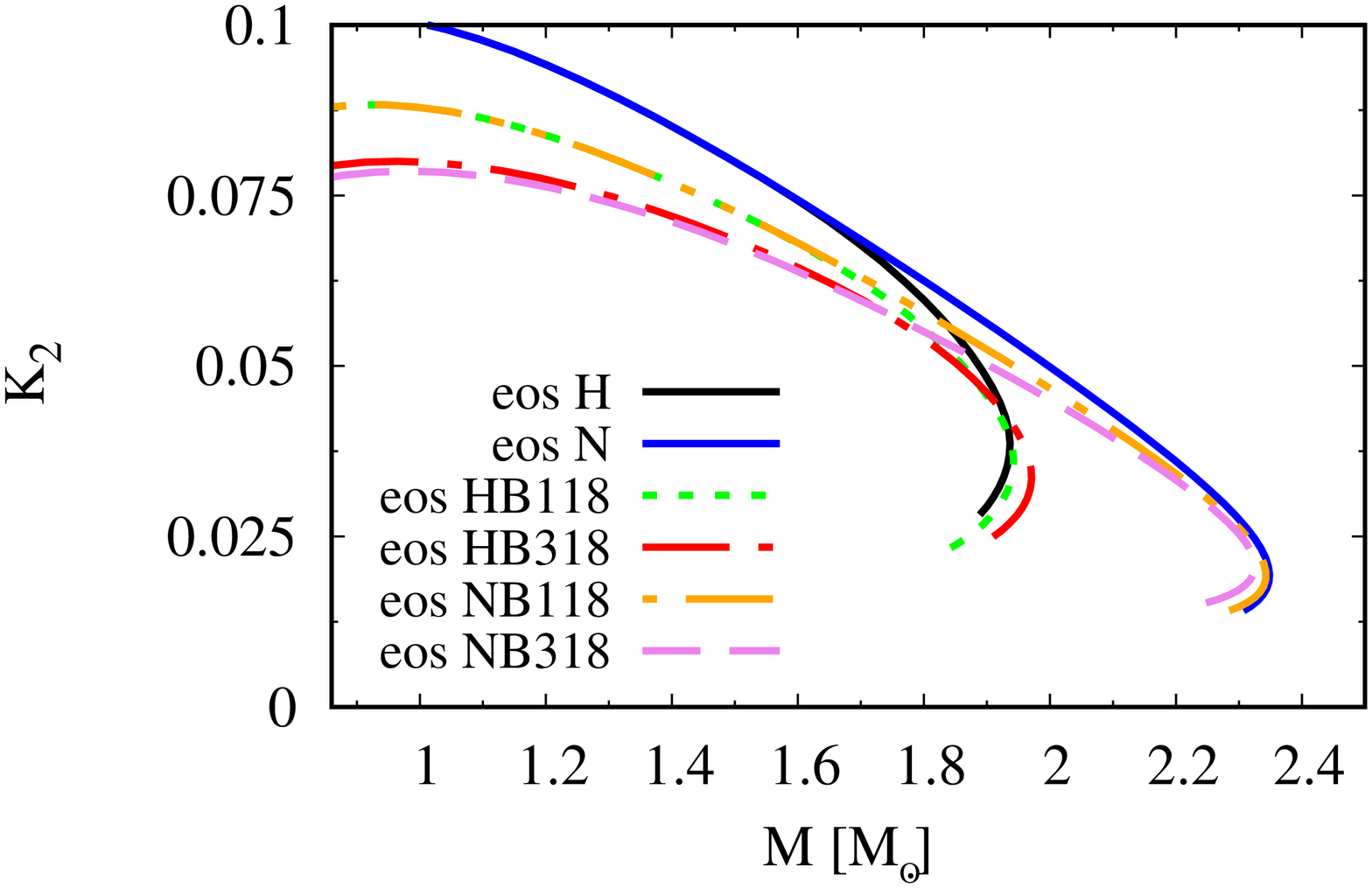}}\\
\subfloat[]{\includegraphics[angle=-90,width=0.5\textwidth]{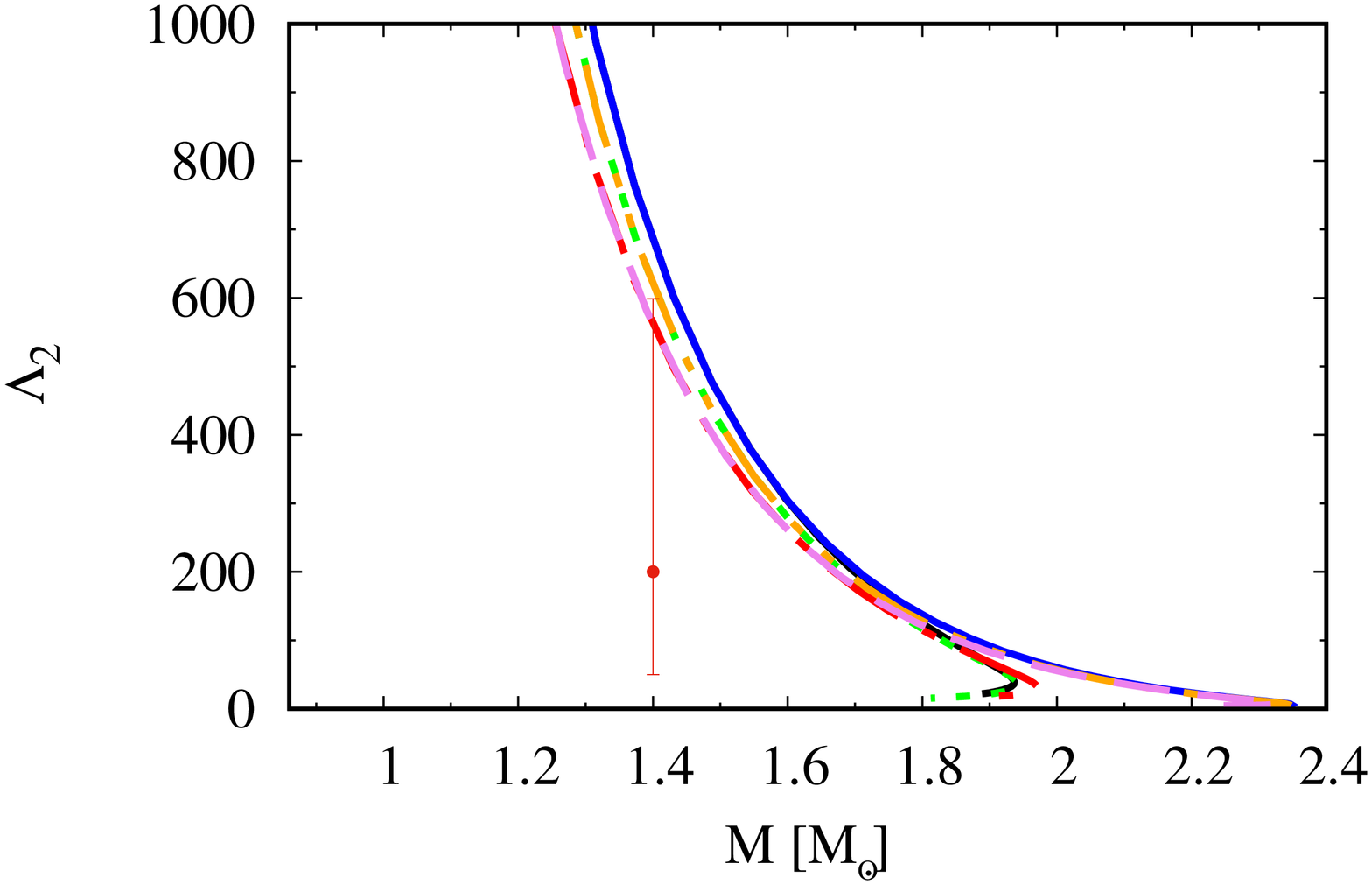}}
\quad
\subfloat[]{\includegraphics[angle=-90,width=0.5\textwidth]{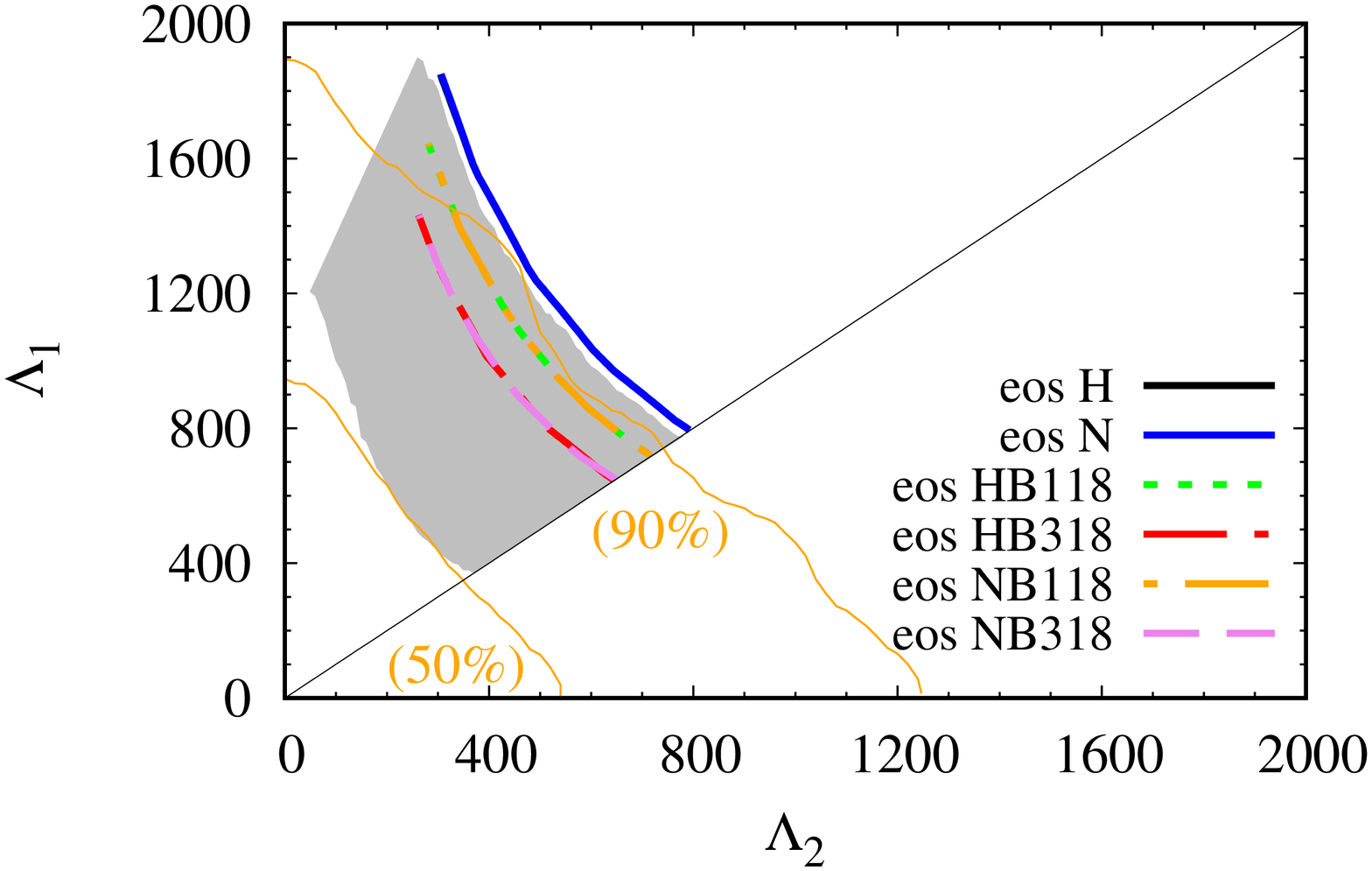}}
\caption{Top figures: Love number  as a function of a) the compactness and b) of the stellar mass. Bottom figures: Dimensionless tidal polarizability c) as a function of the stellar mass and d) $(\Lambda_1,\Lambda_2)$ window obtained from the LIGO and Virgo collaboration.}
\label{fig3}
\end{figure}

In Fig. \ref{fig3}(a) and Fig.\ref{fig3}(b) the Love number is plotted as a function of the compactness and stellar mass and in Fig. \ref{fig3}(c) the dimensionless tidal polarizability is shown as a function of the stellar mass.
If we compare Fig. \ref{fig3}(a) with the ones produced with other models in the literature, we observe that the second Love number lies at about the same range as many of the results obtained with other RMF parametrizations \cite{PhysRevC.99.045202} (notice the difference in the x scale), but below most of the results found with different versions of the quark-meson coupling model \cite{QMC2019}. These results are clearly model dependent, but our point here is to confirm that most of the differences reside on the constitution of the star (containing hyperons or not) and the effects of the magnetic field are noticeable, but minor. This observed features are certainly expected because in our choice of modelling the magnetic field, its value at the crust is set to be $B=10^{15}$ G and this low magnetic field hardly affects the EOS.  
In Fig. \ref{fig3}(c), $\Lambda$ is plotted alongside recent results of the canonical $\Lambda_{1.4} = 190^{+390}_{-120}$ obtained by the LIGO and Virgo Collaboration \cite{PhysRevLett.121.161101} and we see that our results barely touch the error bar.
Again, it is worth pointing out that a similar result is obtained in \cite{QMC2019}
and it is due to the choice of parameters. More experimental results should be obtained before any strong conclusion can be drawn. 

In Fig. \ref{fig3}(d) we show the tidal deformabilities $(\Lambda_1,\Lambda_2)$ for the binary system ($m_1,m_2$), with $m_1 > m_2$. The plots are calculated using the equation for the chirp mass
\begin{equation}
M_{chirp} = (m_{1}m_{2})^{3/5}(m_{1}+m_{2})^{-1/5},
\end{equation}
and the diagonal solid line corresponds to the case $m_1 = m_2$. The lower and upper solid orange lines correspond to 50$\%$ and 90$\%$ confidence limits respectively, which are obtained from the GW170817 event. The shadow region represents recently published theoretical results \cite{PhysRevC.99.045202}, obtained with non-magnetized EOS. We see, once again, that most of our results lie within the confidence limits and the effects of the magnetic field are strong enough to make our results in agreement with the experimental region values. It is worth mentioning that we do not expect such high magnetic fields to be present in most of observed stars. 
Nevertheless, we can see that all the results presented in Figures \ref{fig3}(a), \ref{fig3}(b) and \ref{fig3}(c) are sensitive to the presence of hyperons and are  affected by the intensity of the magnetic field, at least for less massive stars. 
%

To understand better our results, we display both the deformability and the Love number for the maximum mass and the canonical stars in Table \ref{TL3}. As we can see, strong magnetic fields can reduce the deformability $\Lambda$ by almost 19$\%$. The influence of the magnetic field in the deformability is way bigger than in the radius, whose reduction is less than $1\%$. This can indicate that measurements of the $\Lambda$ can put a strong constraint in the EOS. The same can be said about the Love number $k_2$ whose reduction was about $17\%$. It is worth noticing that for the canonical mass, there are no hyperons in the core. The small difference between $\Lambda$ for  NB318 and HB318 is due to different values of the $\epsilon_c$ at Eq.~(\ref{EL5}). There is also a difference of $21\%$ in the maximum mass due to the effect of the strong magnetic field. However, we have to keep in mind that the these maximum masses are not identical.

\begin{table}[ht]
\begin{center}
\caption{Deformabilities and Love numbers for the maximum and the canonical star.} 
\label{TL3}
\begin{tabular}{|c|c|c||c|c|}
\hline
  EOS   &  $\Lambda$ ($M_{max})$ & $k_2~(M_{max})$  & $\Lambda$ ($1.4M_\odot$) & $k_2~(1.4M_\odot$)      \\
  \hline
  N & 5.99    & 0.0193 & 684 &   0.0851\\
 \hline
 NB118 & 6.52 & 0.0198 & 621 &  0.0768
 \\
\hline
 NB318 & 7.38 &  0.0205 & 559 & 0.0711   \\
\hline
 \hline
  H & 39.47   &  0.0387 & 684 & 0.0851 \\
 \hline
 HB118 & 
 36.67 & 0.0363 & 621 & 0.0766  \\
\hline
 HB318 & 30.57 & 0.0334  & 564 &   0.0719 \\
\hline
\hline
\end{tabular} 
\end{center}
\end{table}

A more recent constraint concerns the radii of the canonical stars, the ones with $M=1.4  M_\odot$. Although in the past, studies pointed that the radii of the canonical stars could be as larger as 17 km \cite{17km}, nowadays this value is believed to be significant lower. More conservative results point towards a maximum radius of 13.9 km \cite{Hebeler,2013ApJ...773...11H}, while more radical studies point to 13 km as the maximum radius \cite{Lattimer2013,Lattimer2014}. Recently, the LIGO and Virgo collaboration stated that the tidal polarizability of canonical stars should lie in the range $70 \leq \Lambda_{1.4} \leq 580$ \cite{PhysRevLett.121.161101} and this restriction imposed another constraint to the radii of the corresponding stars. 
According to \cite{PhysRevC.98.035804}, the values should lie in the region $11.82$ km $ \leq R_{1.4 M_{\odot}} \leq 13.72$ km and
according to \cite{PhysRevLett.121.161101}, in the range $10.5$ km $ \leq R_{1.4 M_{\odot}} \leq 13.4$ km.  Whichever constraint we consider correct, we see that our results for the radii are very close to the border of these ranges. Nevertheless, a very new result 
indicates that the canonical neutron star radius cannot excceed 11.9 km~\cite{Capano2020}. If it turns to be confirmed, this could imply in a revision of the known EOS or the gravity theory itself,
as done in \cite{PhysRevD.100.024043}, for instance.

\subsection{Lower mass limit}

Now, let's take a closer look at the other edge of the neutron star family  - the low mass neutron stars. The minimum stable neutron star is about 0.1 $M_\odot$, although a more realistic minimum stems from the neutron star origin in a supernova. Lepton-rich proto-neutron stars are unbound if their masses are less than about 1.0 \cite{Lattimer2004,Haensel2002AA} $M_\odot$. We show here how strong magnetic field affects neutron stars with masses lying from 1.0 to 1.25 solar masses. At such low value no hyperon is present, so we are only dealing with nucleonic neutron stars.  

We plot in Table \ref{TL4} the radii, the tidal deformability $\Lambda$, the Love number $k_2$ and central densities of these stars for different values of magnetic field.

\begin{table}[ht]
\begin{center}
\caption{Properties of low mass neutron stars for different values of magnetic field.} 
\label{TL4}
\begin{tabular}{|c|c|c|c|c|c|c|c|}
\hline
  EOS   &  $M$ ($M_\odot)$ & R $(km)$  & $\Lambda$ & $k_2$ & $\epsilon_c~(fm^{-4})$  & $B_c$  (G)    \\
  \hline
  N & 1.04   & 13.52 & 3361 &  0.0993 & 1.34 & - \\
 \hline
 NB118 & 1.04 & 13.53 & 2880 & 0.0873 & 1.33 & 1.7 x 10$^{16}$ \\
\hline
 NB318 & 1.04 & 13.38 & 2552 & 0.0796 &  1.31 & 5.1 x 10$^{16}$ \\
\hline
 \hline
  N & 1.14   & 13.56 & 2028 & 0.0961 &  1.42 & -\\
 \hline
 NB118 & 1.14 & 13.56 & 1844 & 0.0850 &   1.42 &  2.0 x 10$^{16}$ \\
\hline
 NB318 & 1.14 & 13.50 & 1628 & 0.0784  & 1.40 &  6.3 x 10$^{16}$    \\
\hline
 \hline
  N & 1.25   & 13.60 &  1288  & 0.0922 &   1.52 & -  \\
 \hline
 NB118 & 1.25 & 13.61 & 1156 & 0.0823 &  1.52 &  2.5 x 10$^{16}$  \\
\hline
 NB318 & 1.25 & 13.56 & 1018 & 0.0762  & 1.49 &  7.6 x 10$^{16}$ \\
\hline
\end{tabular} 
\end{center}
\end{table}

When compared with the canonical mass, the influence of strong magnetic field in neutron star radii is about three times stronger than for the low mass edge. For a 1.04 M$_\odot$ the radius drops from 13.52 km (non-magnetized stars) to 13.38 km ($B_0 = 3.0~\times~10^{18}$ G). A reduction of 0.14 km against only 0.05 km in the canonical mass star. As we increase the mass, the influence of the magnetic field becomes smaller even if the strength of the magnetic field becomes higher, according to 
eq. (\ref{EL5}), and the differences in the radii are 0.06 km and 0.04 km to 1.14$M_\odot$ and 1.25$M_\odot$, while the maximum central magnetic field (for $B_0 = 3.0~\times~10^{18}$ G) are: 5.1, 6.3 and 7.6  x 10$^{16}$ G respectively. This indicates that the magnetic field plays a more important role at the low density limit than at the high one. A magnetic field around 5 x 10$^{16}$ G is enough to produce effects on low mass stars, while a 50 times stronger field does not affect massive stars.

While for the radii the bigger difference is about $1.1\%$, for the deformability this difference achieves $24\%$, as its value drops from 3361 to 2552 for a strong magnetic field. For the Love number $k_2$ the difference can reach $21\%$. 
Our results imply that strong magnetic field plays an important role in the deformability of the neutron stars, specially for the  low masses one,  with possible consequences on the interpretation of the detected gravitational waves signatures.

\section{ Neutron star oscillations}

Oscillations in neutron stars can be excited by the violent dynamics of the binary system. The theory to study the quasi-normal modes of compact stars is well established \cite{1967ApJ...149..591T,1968ApJ...152..673T,1983ApJS...53...73L}. In this work we use the Lindblom and Detweiler method that is widely used to compute the fluid modes. In this section we briefly present the perturbative formalism generally used to compute neutron star oscillations.

Before to present the non radial oscillations equations, we would like to discuss why we use spherical symmetry in the treatment of Einstein's Equations. This is a very important issue when we study the tidal deformability and QNM formalism in neutron stars. As we explained in the discussion corresponding to the magnetized equation of state, the monopole nature of the chaotic magnetic field inside the star validates the use of the spherical symmetry for the background metric. On the other hand, it is well known that next to the star surface, the  magnetic field intensity has lower values than in the star inner regions. Therefore its  energy is considerably smaller than the gravitational field energy \cite{Sotani2007}, for this reason the backreaction of the magnetic field on the metric  can be neglected, as seen in the equation below (which can be used next to the surface of the star):
\begin{equation}
    \frac{B^2}{4\pi<\rho_0 >c^2} \simeq 2.2 \times 10^{-7} \left( \frac{B}{10^{15}G}\right)^2
    \left( \frac{ 1.4 M_{\odot}}{ M } \right)
    \left( \frac{R}{15 Km}  \right)^3
\end{equation}
where $<\rho_0>$ is the mean density of the star. Therefore if we can neglect the magnetic field backreaction near the surface of the star, it is reasonable to neglect the magnetic field reaction outside it. Then in the deduction of the tidal deformability and QNM equations, it is plausible to assume a spherical symmetric background metric inside and outside the star.

After  the polar non-radial perturbations of a non-rotating star can be described through a set of equations presented in \cite{1983ApJS...53...73L,1985ApJ...292...12D}.  The perturbed metric tensor  reads
\begin{eqnarray}
ds^2 & = & -e^{\nu}(1+ h_{1})dt^2  - h_{2}dtdr   
+ e^{\lambda}(1 - h_{3})dr^2 \nonumber \\
&& + r^2(1 - h_{4})(d\theta^2 + \sin^2\theta d\phi^2),
\end{eqnarray}
where the metric perturbations are given by
\begin{eqnarray}
h_{1} & = & r^{\ell}H_0Y^{\ell}_{m}e^{i\omega t}   \\
h_{2} & = & 2i\omega r^{\ell+1}H_1Y^{\ell}_me^{i\omega t}\\ 
h_{3} & = & r^{\ell}H_0Y^{\ell}_{m}e^{i \omega t} \\ 
h_{4} & = & r^{\ell}KY^{\ell}_{m}e^{i \omega t} \\ \nonumber
\end{eqnarray}
and the polar perturbations in the fluid are given  by the following Lagrangian displacements
\begin{eqnarray}
\xi^{r} &=& r^{\ell-1}e^{-\lambda/2}WY^{\ell}_{m}e^{i\omega t}, \\
\xi^{\theta} &=& -r^{\ell - 2}V\partial_{\theta}Y^{\ell}_{m}e^{i\omega t}, \\
\xi^{\phi} &=& -r^{\ell}(r \sin \theta)^{-2}V\partial_{\phi}Y^{\ell}_{m}e^{i\omega t} ,
\end{eqnarray}
where $Y^{\ell}_{m}(\theta,\phi)$ are the spherical harmonics, and $l$ is restricted to the $l = 2$ component, which dominates the emission of gravitational waves. 

Non-radial oscillations are then described by the following set of first order linear differential equations \cite{1985ApJ...292...12D}:
\begin{eqnarray}
H_1' &=&  -r^{-1} [ \ell+1+ 2Me^{\lambda}/r +4\pi   r^2 e^{\lambda}(p-\epsilon)] H_{1}  
 +  e^{\lambda}r^{-1}  \left[ H_0 + K - 16\pi(\epsilon+p)V \right] ,      \label{osc_eq_1}  \\ \nonumber \\
 K' &=&    r^{-1} H_0 + \frac{\ell(\ell+1)}{2r} H_1   - \left[ \frac{(\ell+1)}{r}  - \frac{\nu'}{2} \right] K  
 - 8\pi(\epsilon+p) e^{\lambda/2}r^{-1} W \:,  \label{osc_eq_2} \\ \nonumber \\
 W' &=&  - (\ell+1)r^{-1} W   + r e^{\lambda/2} [ e^{-\nu/2} \gamma^{-1}p^{-1} X
 - \ell(\ell+1)r^{-2} V + \tfrac{1}{2}H_0 + K ] \:,  \label{osc_eq_3}
\end{eqnarray}

\begin{eqnarray}
X' &=&  - \ell r^{-1} X + \frac{(\epsilon+p)e^{\nu/2}}{2}  \Bigg[ \left( r^{-1}+{\nu'}/{2} \right)H_{0} 
+ \left(r\omega^2e^{-\nu} + \frac{\ell(\ell+1)}{2r} \right) H_1    + \left(\tfrac{3}{2}\nu' - r^{-1}\right) K   \nonumber  \\
&&    - \ell(\ell+1)r^{-2}\nu' V  -  2 r^{-1}   \Biggl( 4\pi(\epsilon+p)e^{\lambda/2} 
 + \omega^2e^{\lambda/2-\nu}  - \frac{r^2}{2}  (e^{-\lambda/2}r^{-2}\nu')' \Biggr) W \Bigg]  \:,
\label{osc_eq_4}
\end{eqnarray}
where the prime denotes a derivative with respect to $r$ and  $\gamma$ is the adiabatic index. The function $X$ is given by
\begin{eqnarray}
X =  \omega^2(\epsilon+p)e^{-\nu/2}V - \frac{p'}{r}e^{(\nu-\lambda)/2}W   \nonumber \\
  + \tfrac{1}{2}(\epsilon+p)e^{\nu/2}H_0 ,
\end{eqnarray}
and $H_{0}$ fulfills the algebraic relation 
\begin{eqnarray}
H_{0}= \dfrac{1}{b_1}(b_2  X -  b_3 H_{1}  + b_4 K),
\end{eqnarray}
with
\begin{eqnarray}
b_1 &=&  3M + \tfrac{1}{2}(l+2)(l-1)r + 4\pi r^{3}p  ,  \\
b_2 &=&  8\pi r^{3}e^{-\nu /2}   , \\
b_3 &=&  \tfrac{1}{2}l(l+1)(M+4\pi r^{3}p)-\omega^2 r^{3}e^{-(\lambda+\nu)}  ,  \\
b_4 &=&  \tfrac{1}{2}(l+2)(l-1)r - \omega^{2} r^{3}e^{-\nu}  \nonumber \\
    &&    -r^{-1}e^{\lambda}(M+4\pi r^{3}p)(3M - r + 4\pi r^{3}p)   .
\end{eqnarray}
Outside the star, the perturbation functions that describe the motion of the fluid vanish and the  system of differential equations reduces to the Zerilli equation:
\begin{equation}
\frac{d^{2}Z}{dr^{*2}}=[V_{Z}(r^{*})-\omega^{2}]Z,
\end{equation}
where $Z(r^{*})$ and $dZ(r^{*})/dr^{*}$  are related to the metric perturbations  $H_{0}(r)$ and $K(r)$ by transformations given in Refs. \cite{1983ApJS...53...73L,1985ApJ...292...12D}.  The ``tortoise'' coordinate is $r^{*} = r + 2 M \ln (r/ (2M) -1)$, and  the effective potential  $V_{Z}(r^{*})$ is given by
\begin{eqnarray}
V_{Z}(r^{*}) = \frac{(1-2M/r)}{r^{3}(nr + 3M)^{2}}[2n^{2}(n+1)r^{3} + 6n^{2}Mr^{2}  \nonumber \\
 + 18nM^{2}r + 18M^{3}],
\end{eqnarray}
with $n= (l-1) (l+2) / 2$. 

The quasi-normal modes have to be determined by a two stage process, one inside the star and the other outside. Inside the star we have to obtain the coefficients of the differential equations, which are defined at each point. Those coefficients depend directly on the mass, metric, pressure, energy density, etc., and these quantities can be obtained from the stellar structure equations. The integration
outside the star is performed with the use of the Zerilli equations. All the procedure has to respect boundary conditions at the centre, surface of the star and at infinity. 
All the equations are numerically integrated for the quadrupole oscillations $(l=2)$. More details about the method can be find in Ref. \cite{2017PhRvC..95b5808F}. This procedure allows us to obtain $\omega$ for each value of the central density of the star, or equivalently for each value of the stellar mass.  The real part of $\omega$  is the pulsation frequency ($f = \mathrm{Re}(\omega)/2 \pi$) and the imaginary part is the inverse of the damping time of the mode due to the gravitational wave emission ($\tau= 1/\mathrm{Im}(\omega)$).

\begin{figure}[ht]
\subfloat[]
{\includegraphics[angle=-90,width=0.5\textwidth]{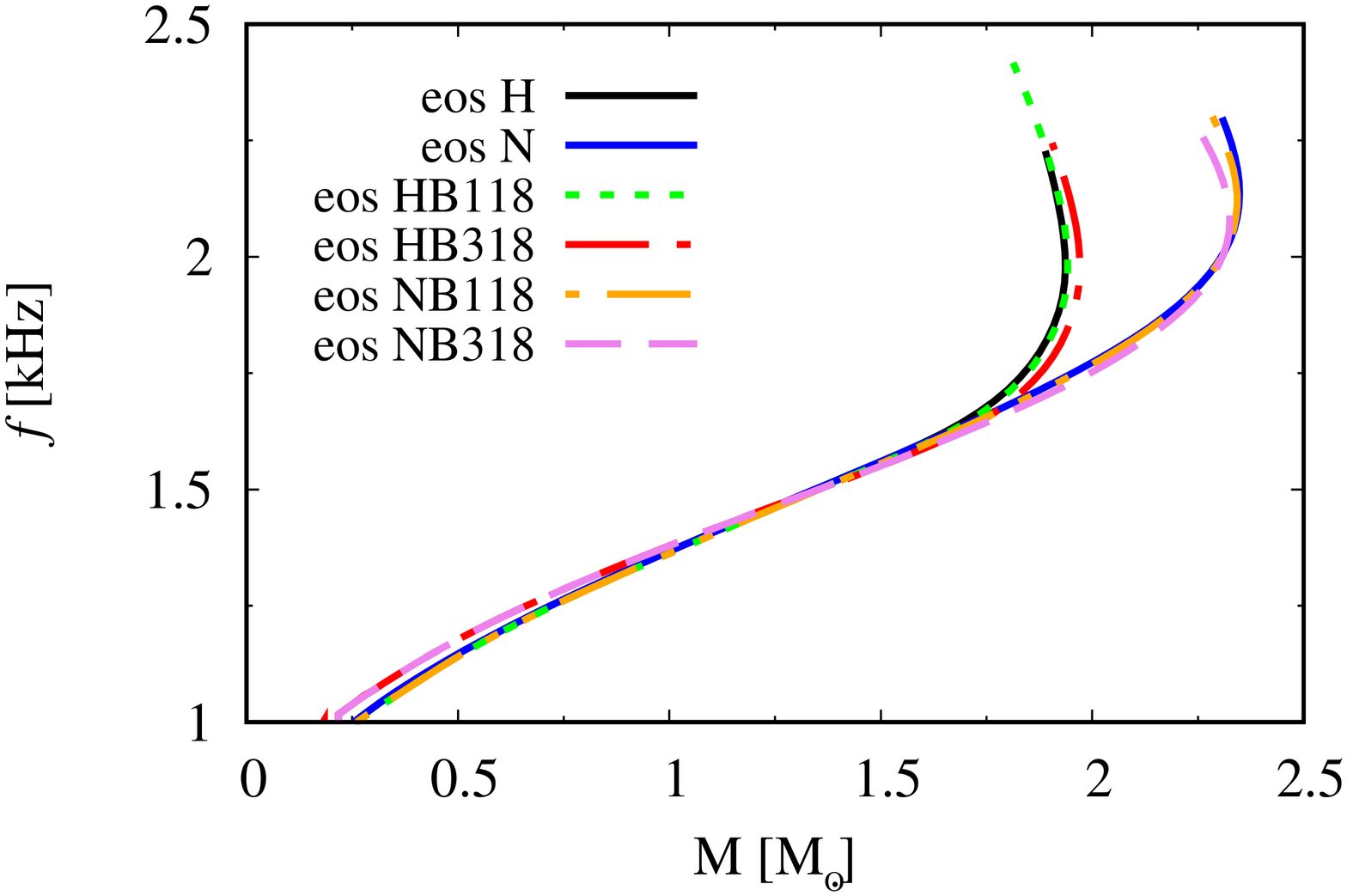}}
\subfloat[]
{\includegraphics[angle=-90,width=0.5\textwidth]{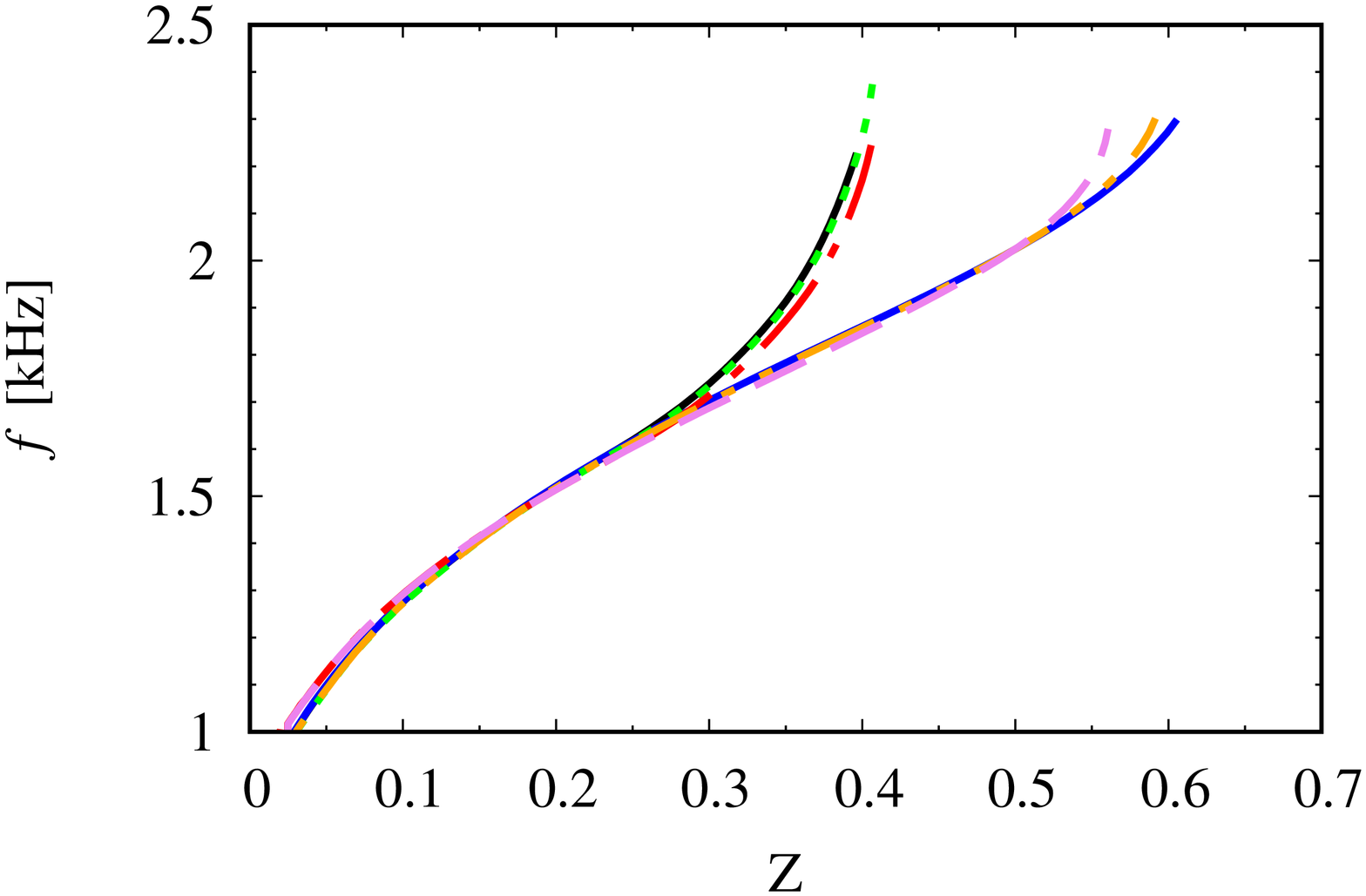}}\\
\subfloat[]
{\includegraphics[angle=-90,width=0.5\textwidth]{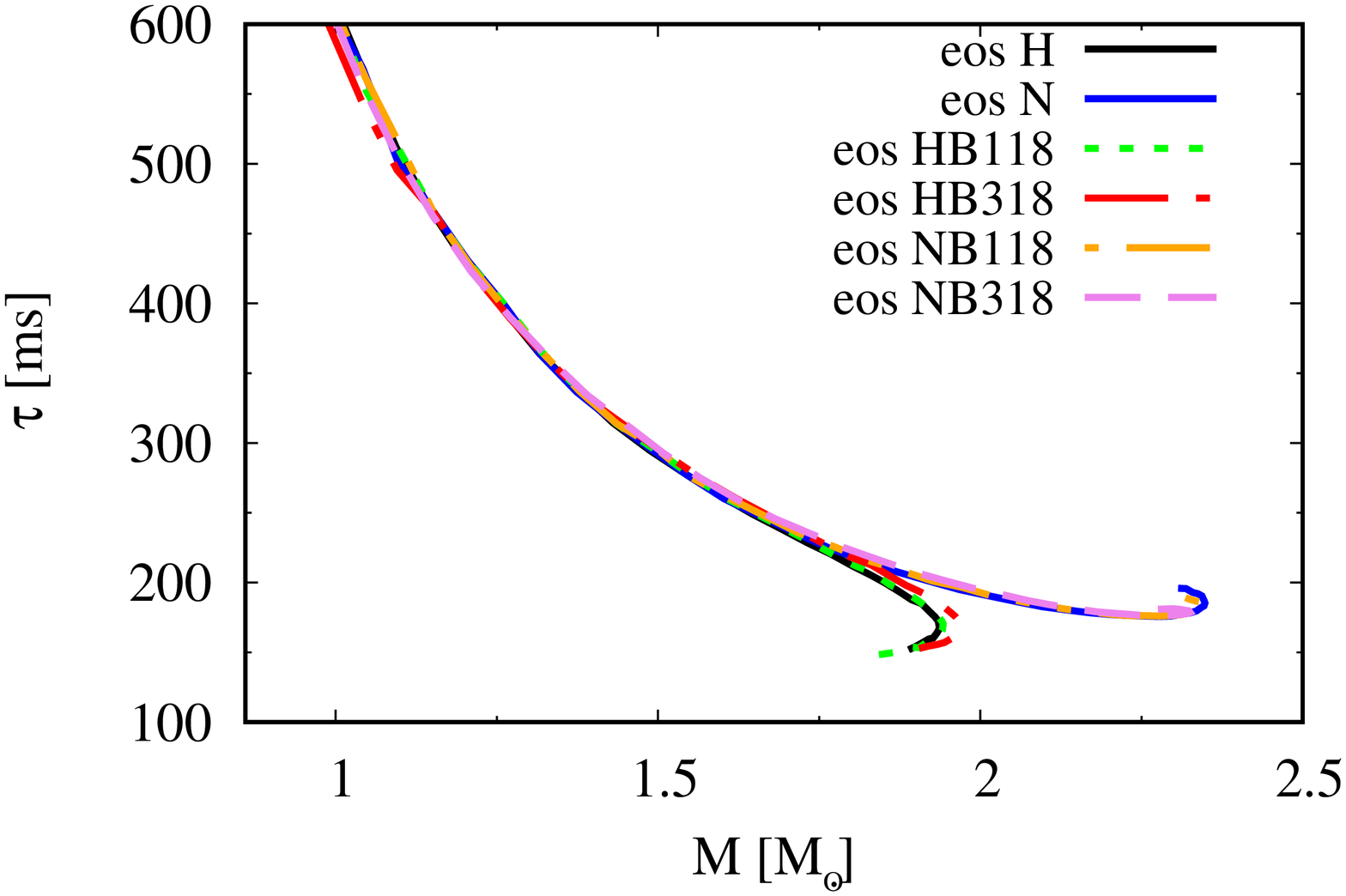}}
\subfloat[]
{\includegraphics[angle=-90,width=0.5\textwidth]{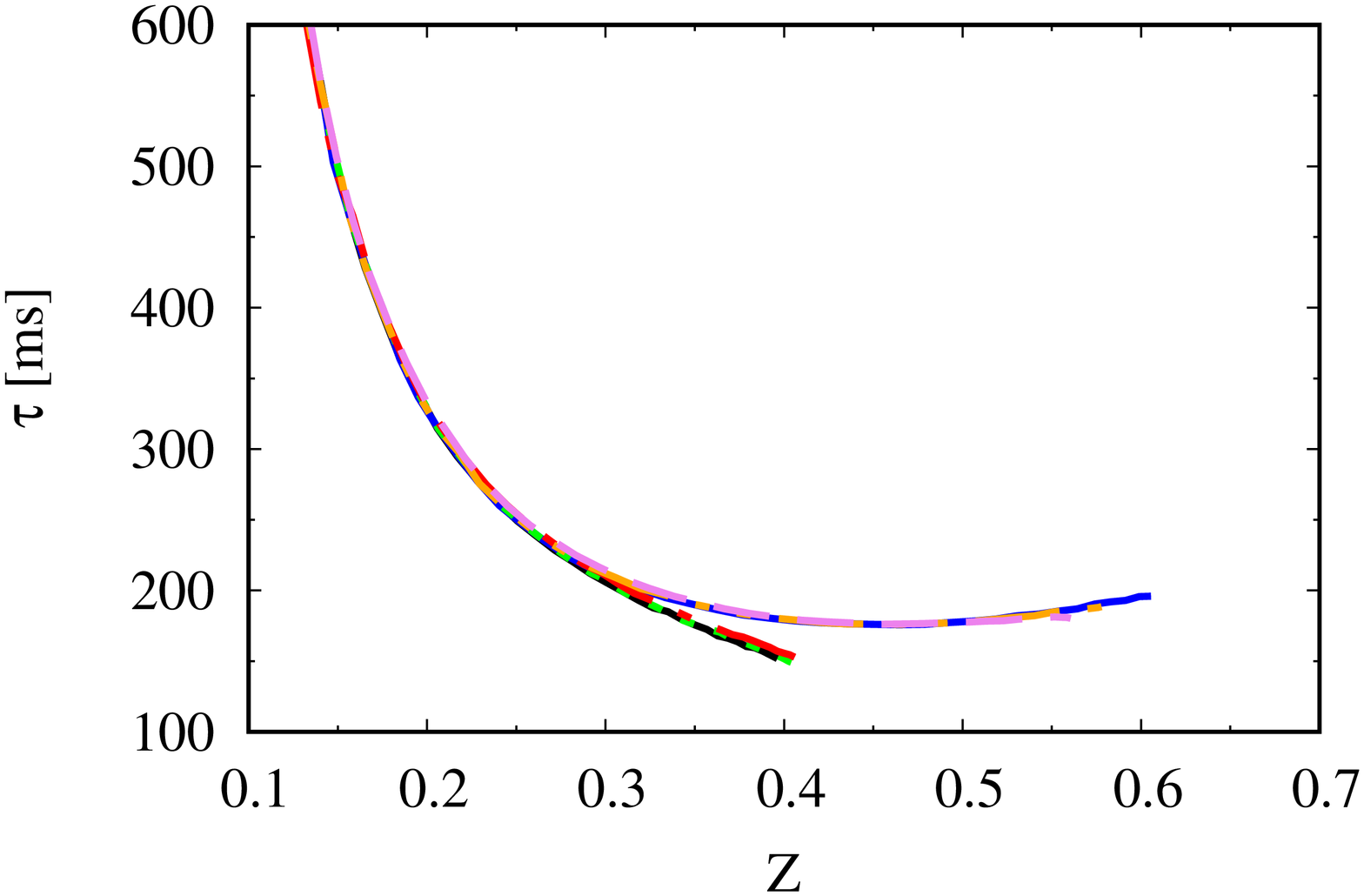}}\\
\caption{Top figures: Fundamental mode frequency  as a function of a) the stellar mass and b) of the redshift. Bottom figures: Damping time c) as a function of the stellar mass and d) of the redshift.}
\label{fig4}
\end{figure}

As stated in the Introduction, we only focus only on the $f$-mode because it is easily excited in astrophysical events while it is expected to be detected by third generation detectors in the near future. In the next lines we present and discuss our main results.
In Fig. \ref{fig4} (a)-(b) we show the plots for the frequency of the quadrupole fundamental fluid mode as a function of the stellar mass and redshift, from where we observe that only the magnetic field with $B_0$ =  3$\times 10^{18}$G produces a small effect on the frequency (not always obvious from the figures) and this result can be noted for both types of stars, with and without hyperons. This effect is related to the small increase in mass obtained with highly magnetized matter and this increase is larger for hyperonic than for nucleonic matter, as seen in Table \ref{TL2}.

The same qualitative behaviour is observed for the frequency as a function of the redshift. On the other hand, it is clear that the constitution of the star plays a very important role and the frequency generated by massive hyperonic stars (larger than 1.8 $M_\odot$) is greatly increased as compared with their nucleonic counterparts. It is also observed that the gravitational wave frequency of the fundamental mode for our models fall in the range of 1.4 - 2 kHz for stars with masses between 1.4 - 2.4 M$_{\odot}$, this values corresponds with previous results in the literature  \cite{2017PhRvC..95b5808F,Flores_2018} obtained with less realistic EOS. We can see that in general a high magnetic field produce little effect on the frequency window.

\begin{table}[ht]
\begin{center}
\caption{Frequency and damping time for the maximum mass and the canonical star.} 
\label{TL5}
\begin{tabular}{|c|c|c||c|c|}
\hline
  EOS   &  $f$ ($M_{max}$) & $\tau~(M_{max})$  & $f$ ($1.4M_\odot$) & $\tau~(1.4M_\odot$)      \\
  \hline
  N &  2.13489  & 185.364 & 1.51173 & 336.6089  \\
 \hline
 NB118 & 2.10878 & 182.074 & 1.50948 & 336.7872   \\
\hline
 NB318 & 2.08084 &  178.495 &  1.51260   &  334.2141  \\
\hline
 \hline
  H &   1.97587  & 167.886 & 1.51126   &  339.2875\\
 \hline
 HB118 & 1.97332    & 170.4868 &1.50900   &   338.3175 \\
\hline
 HB318 & 1.99720  & 167.0384 & 1.51303
  &  331.4330 \\
\hline
\end{tabular} 
\end{center}
\end{table}

 We also present, in Fig. \ref{fig4}(c)-(d) the plots of the damping time as a function of the mass and gravitational redshift. The damping time for a typical 1.4 M$_{\odot}$ neutron star is near 350 ms, as previously obtained in \cite{2017PhRvC..95b5808F} for {\it strange stars} and a high magnetic field does not affect this value. Once again, we observe that only the strongest magnetic field produces non-negligible effects, better noticed in hyperonic stars. As can be seen in Fig. \ref{fig4}(c), the presence of hyperons produces a decrease in the damping time for stars with masses beyond (to the left on the mass-radius diagram) the maximum mass, but those stars are not expected to be stable. 
 
 In Table~\ref{TL5} we plot the fundamental mode frequencies and the damping time for the canonical and the maximum mass stars obtained with different magnetic fields. Unlike the deformability $\Lambda$, the frequencies and time damping do not vary significantly, even for strong magnetic fields. 

\subsection{Lower mass limit}

For the sake of completeness, we also plot the fundamental mode frequencies and the time damping for the low mass limit in Tab~\ref{TL6}. As we can see, even in the low mass limit there is no significant modification in the frequencies and time dumping due to the magnetic field. 

\begin{table}[ht]
\begin{center}
\caption{Properties of low mass neutron stars for different values of the magnetic field.} 
\label{TL6}
\begin{tabular}{|c|c|c|c|c|}
\hline
  EOS   &  $M$ ($M_\odot)$ & $f$ $(kHz)$  & $\tau~(ms)$    \\
  \hline
  N & 1.04   &  1.3840   &  568.0611  \\
 \hline
 NB118 & 1.04 & 1.3813  & 560.2122 \\
\hline
 NB318 & 1.04 & 1.3927   &  559.8997   \\
\hline
 \hline
  N & 1.14   & 1.4257   &  476.1311  \\
 \hline
 NB118 & 1.14 &  1.4174  &  476.9140 \\
\hline
 NB318 & 1.14 & 1.4284   &  463.3985  \\
\hline
 \hline
  N & 1.25   &  1.4644   &  400.8562 \\
 \hline
 NB118 & 1.25 & 1.4602  &   400.0862  \\
\hline
 NB318 & 1.25 & 1.4657 & 396.0027  \\
\hline
\end{tabular} 
\end{center}
\end{table}

\subsection{On the amplitude and detectability of the gravitational wave signal}

 It is well known that, in the QNM formalism the gravitational signal has the form:
\begin{equation}
h(t) = \textit{h}e^{-t/\tau}{\rm sin}[2\pi f t]
\end{equation}
 where $\textit{h}$ is the amplitude , \textit{f} is the   fundamental mode frequency, $\tau$ is the damping time and $h$ is given by
\begin{equation}
    \textit{h} \sim 2.4 \times 10^{-20}\left( \frac{E_{gw}}{10^{-6}M_{\odot}c^2}    \right)^{1/2}
    \left( \frac{10 {\rm kpc} }{d} \right)
    \left( \frac{1 {\rm kHz}}{f} \right)
    \left( \frac{{1 \rm ms}}{\tau}  \right)^{1/2}
    \label{detection1}
\end{equation}
 where $E_{gw}$ is the energy released trough the fundamental mode and $d$ is the distance to the source \cite{Echevarria1989,Apostolatos2001}.

 The signal-to-noise ratio at the detector reads \cite{Echevarria1989,Apostolatos2001}
\begin{equation}
\left( \frac{S}{N} \right)^2 = \frac{4 Q_{\textit{F}}^2}{1+4 Q_{\textit{F}}^2} \; \frac{ h^2 \tau}{2 S_n} ,
\label{detection2}
\end{equation}
 where $Q_{\textit{F}}^2 \equiv \pi f \tau$ is the quality factor and $S_n$ is the noise power spectral density.

 From Eqs. \eqref{detection1} and \eqref{detection2} we can obtain:
\begin{equation}
 \left( \frac{E_{\rm gw}}{M_{\odot} c^2}  \right)  = 3.47 \times 10^{36} \left( \frac{S}{N} \right)^2   \frac{1+4 Q_{\textit{F}}^2}{4 Q_{\textit{F}}^2}  \left( \frac{d}{10 {\rm kpc} }  \right)^2 \left( \frac{f}{1 {\rm kHz}}  \right)^2 \left( \frac{S_n}{1 {\rm Hz^{-1}} } \right) .
\label{detection3}
\end{equation}

 We'll consider two detectors: the first with $S_n^{1/2} \sim 2 \times 10^{-23} \, \mathrm{Hz}^{-1/2}$ (Advanced LIGO/VIRGO) at $\sim$kHz \cite{abbott2017b}, and the second one with $S_n^{1/2} \sim 10^{-24} \, \mathrm{Hz}^{-1/2}$ ( Einstein telescope) at the same frequencies \cite{Einstein}. We now consider that gravitational radiation of a magnetar could be observed in our galaxy ($d \sim 10$ kpc). If the star has a mass of 2.0 M$_\odot$, $f=1.75$ kHz and $\tau=$200 ms, then we can see that $E_{gw}>2.67\times 10^{-10}M_{\odot}c
^2$ for a (S/N)$>5$ at the Einstein telescope and $E_{gw}>1.06\times 10^{-7}M_{\odot}c
^2$ for a (S/N)$>5$ at Advanced LIGO/VIRGO.

\section{Summary and Final Remarks}

In the present work we have analysed the influence of strong magnetic fields on equations of state that describe both nucleonic and hyperonic matter and the resulting effects on the Love number, tidal polarizabilities, stellar radii and the fundamental quasi-normal oscillation mode. To compute the EOS, the chaotic field approximation has been used and an energy density dependent magnetic field prescription utilized so that the magnetic field of the crust never exceeds the observed $10^{15}$ G. 
We have seen that the constitution of the stars (nucleonic or hyperonic) plays an important role in the computation of the Love number and the tidal polarizability. Very large magnetic fields (of the order of $3 \times 10^{18}$ G are present only in very massive stars). We see that even if magnetic fields are around 50 - 60 times higher than those found in lower mass stars, their effects are not so important. Nevertheless, magnetic fields around 5 x 10$^{16}$ G affects in a significant way low mass neutron stars ($M < 1.4M_\odot$).  While the radii decrease just a little, the deformability $\Lambda$ drops about $19\%$ for the canonical star and more than $23\%$ in the low mass limit. The same can be said about the Love number $k_2$ . Magnetic fields around 10$^{17}$ G are in the limit for stable canonical stars~\cite{Reis}, therefore, neutron stars within this configuration are expected to be very rare in the universe. 

It  is of paramount importance to fully investigate the gravitational wave frequencies of the most important modes of neutron stars because they are expected to be detected in a near future by third  generation  detectors,  like  the  Einstein Telescope. Amongst all the family of modes, we have studied the fundamental mode, which has been the focus of attention for many years, because it has a frequency of nearly 2 kHz and could be detected with an amplitude of $\simeq 10^{-23}$ at 10 Kpc.  For this objective we have also calculated the effect of the magnetic field on the fundamental mode. We have observed that the frequencies practically 
coincide in all cases for stars with masses below $1.8 M_\odot$. However, if more massive stars are considered, the ones constituted by nucleons only present frequencies lower than the ones with  hyperonic cores and this feature might be a way of pointing out the real constituents of neutron stars. 

In all cases, only the strongest magnetic field, i.e., $B_0=3\times 10^{18}$ G alters the frequency. The same behavior is found if we consider the frequency as a function of the redshift. The damping time is typically above 250 ms for masses lower than 1.6 M$_{\odot}$ and for very massive star the damping time is between 100 - 200 ms. We conclude that by the use of the fundamental mode, highly magnetized stars are rare and  could be discriminated only in the limiting case when we use $B_0=3\times 10^{18}$ G, but we reinforce the statement that the different constitutions of the liquid core can be easily tracked.

\begin{acknowledgements}
This work is a part of the project INCT-FNA Proc. No. 464898/2014-5 and it was partially supported by CNPq (Brazil) under grants 301155.2017-8~(DPM), 307932/2017-6 (LBC) and 422755/2018-4 (LBC) and
by Capes-PNPD program (CVF). 
LBC also thanks FAPEMA(Brazil) under grant Universal-01220/18.
\end{acknowledgements}

\bibliographystyle{spphys}       


\end{document}